\documentclass[epj]{svjour}
\usepackage{graphicx,axodraw}
\begin{document}
\title{Landau-Fermi liquid analysis of the  2D $t$-$t'$ Hubbard model}
\author{P. A. Frigeri\inst{1}\fnmsep \thanks{e-mail:
    \texttt{pfrigeri@itp.phys.ethz.ch}} \and C.
  Honerkamp\inst{1,2}\and T. M. Rice\inst{1}}

\institute{ Theoretische Physik,
 ETH-H\"onggerberg, CH-8093 Z\"urich, Switzerland \and
 Department of Physics, Massachusetts
 Institute of Technology, Cambridge MA 02139, USA \thanks{present address}
}

\date{\today}

\abstract{We calculate the Landau interaction function 
$f(\vec{k},\vec{k}')$ for the two-dimensional $t$-$t'$ Hubbard model on the
square lattice using second and higher order perturbation theory. Within the
Landau-Fermi liquid framework we discuss the behavior of spin and charge
susceptibilities as function of the onsite interaction and band filling. In
particular we analyze the role of elastic umklapp processes as driving force for the anisotropic reduction of the compressibility on parts of the Fermi surface.}


\authorrunning{Frigeri, Honerkamp, and Rice} 
\titlerunning{Landau-Fermi liquid analysis of the 2D $t$-$t'$ Hubbard model}
\maketitle
 
\section{Introduction}
%
The  Hubbard model remains one of the most studied
theoretical models in the efforts to understand the high-tem\-perature
superconducting cuprates. Despite their  exceptionally high transition temperatures into
the superconducting state, much of the interest in these materials arises from 
their anomalous normal state properties over a large part of the
tem\-perature-doping phase diagram. These effects occur unequivocally and most
strongly  in the underdoped samples with electron densities close to  half
band filling where the system is a stoichiometric antiferromagnetic Mott insulator. 
In the underdoped regime, the transition into the superconducting state 
occurs out of the so-called pseudogap phase, 
which shows significant deviations from conventional metallic
Fermi-liquid behavior\cite{anderson,timusk}. This  is in 
contrast with sufficiently
overdoped samples where many observations point toward a conventional
transition from a Fermi-liquid into the superconducting state\cite{proust}.
There have been experimental and theoretical studies suggesting that a
quantum critical point hidden underneath the superconducting
dome\cite{tallon,sachdev,chakravarty}  or the onset of Fermi surface
truncation\cite{norman,furukawa,honerkamp} give rise to these phenomena. 
On the other hand one may try to take a more conservative route 
and attempt to understand the
experimental findings within a conventional Fermi-liquid framework. Typically, 
rather specific assumptions on the Landau interaction function 
have to be made\cite{millis} in order to obtain agreement 
with the experimental data.
Thus it appears useful to simply calculate within perturbation theory the Fermi-liquid properties of the
2D Hubbard model and to see whether these assumptions can be justified within
this transparent theoretical framework.

Another motivation for this work is the question whether 
as the density is increased towards half filling 
some kind of drastic change of the low energy properties, 
e.g. corresponding to Fermi surface truncation as indicated by one-loop renormalization group 
calculations\cite{furukawa,honerkamp}, is foreshadowed already 
in a perturbative Fermi liquid picture. Such a partial 
truncation of the Fermi surface has been argued  to be 
seen in angular resolved photoemission
experiments \cite{norman}. 
It is a promising candidate to understand the many anomalies 
of the underdoped normal state of the high-$T_c$ cuprates, as it naturally
combines insulator-like features such as the pseudogap, the observed c-axis
resistivity and the strongly reduced superfluid weight, with residual metallic
properties as evidenced  by the in-plane transport and superconductivity itself.  
Since both the renormalization group 
approaches and the present Fermi liquid analysis are perturbative techniques, 
we will not obtain an accurate description of possible phases
with a truncated Fermi surface,
but we will carefully look for  insulator-like tendencies in these 
calculations as the interaction gets stronger. In particular we analyze 
the influence of umklapp processes between $\vec{k}$-space regions in the
vicinity of the saddle points at $(\pi,0)$ 
and $(0,\pi)$. Elastic umklapp scattering with momentum transfer of
$(\pi, \pi)$ across the Fermi surface is allowed when the Fermi surface 
extends to the Brillouin zone boundary for a certain density range 
close to half filling. 

A Fermi liquid analysis of the 2D Hubbard model close to half filling  
has been presented by Fuseya et
al.\cite{fuseya}. These authors studied the case with only nearest neighbor
hopping, $t$, and found a suppression of the uniform spin susceptibility for
sufficiently strong interaction close to half-filling, which maybe interpreted
as indicating the opening of a spin gap. 
Unfortunately this promising indication turns out to be an
artifact of the calculation scheme used by Fuseya et al. \cite{fuseya}, as
will be discussed below. From renormalization group studies\cite{halboth,honerkamp} 
we know
that nonzero values of the next nearest neighbor hopping, $t'$, can change the
weak coupling physics significantly. Therefore, in this paper we extend the analysis
of the Landau interaction function and related quantities to the case of
$t'\not= 0$.  Particular attention is paid to the anisotropy of the
interaction function for different parts of the Fermi surface.

In the following we first apply a second order perturbative method and discuss the
$\vec{k}$-space integrated and $\vec{k}$-space resolved Fermi liquid
properties of the 2D $t$-$t'$ Hubbard model. Next we describe how ladder summations in the
particle-particle and particle-hole channels, which are necessary to remove
unphysical divergences, modify the tendencies observed
in the second order approach. 
Finally we conclude and compare our results to
other approaches and experimental observations for the high-$T_c$ cuprates.

\section{Second order calculation of the quasiparticle interaction} 
%
The kinetic energy for  the $t$-$t'$ Hubbard model on the 2D square lattice
is given by 
\begin{equation}  
\epsilon_{\vec{k}}=-2t(\cos k_x + \cos k_y )+4t'\cos k_x \cos k_y \, ,  \label{eofk}
\end{equation}  
where $t$ and $t'$ denote the amplitudes for hopping between nearest and next-nearest neighbors, respectively.
The short range Coulomb interaction between the electrons is taken into
account by the usual onsite repulsion 
\begin{equation}  
H_I =  U \sum_{i} n_{i, \uparrow} n_{i, \downarrow} \, .  \label{U}
\end{equation} 
The particle number $N_e$ is controlled by a chemical potential $\mu$.

The Landau interaction function $f_{s,s'}(\vec{k}_F,\vec{k}'_F)$ can be
obtained \cite{theory-of-FL1} from the two-particle vertex 
$\Gamma_{ss'} (\vec{K},\vec{K}',\vec{K}+\vec{Q})$. Here $\vec{K}=(\vec{k},\nu)$
is a short notation for wavevectors and frequencies. $\Gamma$ is a function of
the interaction strength, $U$. To first order in $U$, $\Gamma$ is
wavevector independent, but higher orders lead to a pronounced wavevector dependence.
$\vec{K}$, $\vec{K}'$ and spin indices $s$,$s'$ characterize the two initial particles
and $\vec{K}+\vec{Q}$, $\vec{K}'-\vec{Q}$ and $s$,$s'$ their final
states, and $\vec{Q}=(\vec{q},\omega)$ denotes wavevector and frequency
transfer in the scattering process. The Landau quasiparticle interaction
$f(\vec{k}_F,\vec{k}'_F)$ corresponds to the vertex function in the limit
$\vec{Q} \rightarrow 0$ with $\frac{q}{\omega} \rightarrow 0$ with
$\vec{K}=(\vec{k}_F,0)$ and $\vec{K}'=(\vec{k}'_F,0)$. More precisely  
\begin{equation} \label{rel-vertex-int}  
  f_{s,s'}(\vec{k}_F,\vec{k}'_F)=  z_{k} z_{k'} 
  \lim_{\omega \rightarrow 0 } \lim_{\vec{q} \rightarrow 0} 
  \Gamma_{s,s'}(\vec{K},\vec{K}',\vec{K}+\vec{Q}),  \label{ffromGamma}
\end{equation}
where $\vec{k}_F$, $\vec{k}'_F$ are two wavevectors on the Fermi surface
and $z_k$ is the residue of the pole of the interacting one-particle Green's
function $G(\vec{k}_F,\nu)$, i.e. $z_k=(1-\delta \Sigma / \delta
\nu)^{-1}$ with self-energy $\Sigma(\vec{k},\nu)$.
The two-particle vertex is determined by the coupling function $V_{s
 s'}(\vec{K},\vec{K}',\vec{K}+\vec{Q})=V( \vec{K},s; \vec{K}',s'; \vec{K}+\vec{Q},s)$ via 
{\setlength \arraycolsep{2pt} 
  \begin{eqnarray}  \label{coupling}
    \Gamma_{s,s'}(\vec{K},\vec{K}',\vec{K}+\vec{Q})&=& \delta_{s,s'}[V_{s
      s'}(\vec{K},\vec{K}',\vec{K}+\vec{Q})\nonumber \\
    &-&V_{ss'}(\vec{K}',\vec{K},\vec{K}+\vec{Q})] \nonumber \\
    &+&\delta_{s,-s'}V_{ss'}(\vec{K},\vec{K}',\vec{K}+\vec{Q}),
  \end{eqnarray}  
which expresses the fact that $\Gamma$ has to be antisymmetric with respect to the exchange of two particles with the same spin orientation. 
Then the Landau interaction function can be obtained via
(\ref{coupling}) by taking the  limit (\ref{rel-vertex-int}) in the
perturbation expansion for $V_{ss'}(\vec{K},\vec{K}',\vec{K}+\vec{Q})$. 
The first and second order graphs contributing are shown
in Fig. \ref{graph1-vertex}. 
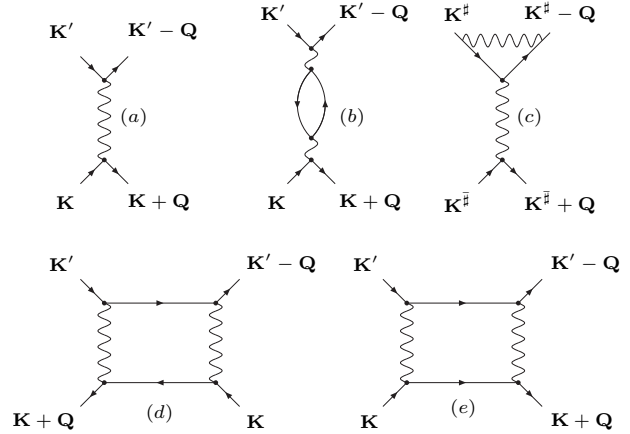
\begin{figure}[h] 
\begin{center} \begin{picture}(370,180)(-20,-100)
\SetScale{0.6}
\ArrowLine(20,20)(35,35)
\Text(10,5)[r]{\scriptsize{$\vec{K}$}}
\ArrowLine(35,35)(50,20)
\Text(30,5)[l]{\scriptsize{$\vec{K}+\vec{Q}$}}
\ArrowLine(20,100)(35,85)
\Text(30,70)[l]{\scriptsize{$\vec{K}'-\vec{Q}$}}
\ArrowLine(35,85)(50,100)
\Text(10,70)[r]{\scriptsize{$\vec{K}'$}}
\Photon(35,35)(35,85){4}{5}
\Text(27,37)[l]{\scriptsize{$(a)$}}
\Vertex(35,35){1.5}
\Vertex(35,85){1.5}
\ArrowLine(270,20)(285,35) 
\Text(160,5)[r]{\scriptsize{$\vec{K}^{\bar \sharp}$}} 
\ArrowLine(285,35)(300,20)
\Text(180,5)[l]{\scriptsize{$\vec{K}^{\bar \sharp}+\vec{Q}$}}
\Photon(260,110)(310,110){4}{5.5}
\ArrowLine(255,115)(285,85)
\Text(180,77)[l]{\scriptsize{$\vec{K}^{\sharp}-\vec{Q}$}}
\ArrowLine(285,85)(315,115)
\Text(160,77)[r]{\scriptsize{$\vec{K}^{\sharp}$}}
\Photon(285,35)(285,85){4}{5}
\Text(177,37)[l]{\scriptsize{$(c)$}}
\Vertex(285,35){1.5}
\Vertex(285,85){1.5}
\ArrowLine(35,-105)(20,-120)
\Text(10,-78)[r]{\scriptsize{$\vec{K}+\vec{Q}$}}
\ArrowLine(120,-120)(105,-105)
\Text(75,-78)[l]{\scriptsize{$\vec{K}$}}
\ArrowLine(105,-55)(120,-40)
\Text(75,-18)[l]{\scriptsize{$\vec{K}'-\vec{Q}$}}
\ArrowLine(20,-40)(35,-55) 
\Text(10,-18)[r]{\scriptsize{$\vec{K}'$}}
\ArrowLine(105,-105)(35,-105)
\ArrowLine(35,-55)(105,-55)
\Text(42,-75)[c]{\scriptsize{$(d)$}}
\Photon(35,-105)(35,-55){4}{5}
\Photon(105,-105)(105,-55){4}{5}
\Vertex(35,-105){1.5}
\Vertex(35,-55){1.5}
\Vertex(105,-105){1.5}
\Vertex(105,-55){1.5}
\ArrowLine(210,-120)(225,-105)
\Text(125,-78)[r]{\scriptsize{$\vec{K}$}}
\ArrowLine(295,-105)(310,-120)
\Text(190,-78)[l]{\scriptsize{$\vec{K}+\vec{Q}$}}
\ArrowLine(295,-55)(310,-40)
\Text(190,-18)[l]{\scriptsize{$\vec{K}'-\vec{Q}$}}
\ArrowLine(210,-40)(225,-55)
\Text(125,-18)[r]{\scriptsize{$\vec{K}'$}}
\ArrowLine(225,-105)(295,-105)
\Text(157,-73)[c]{\scriptsize{$(e)$}}
\ArrowLine(225,-55)(295,-55)
\Photon(225,-105)(225,-55){4}{5}
\Photon(295,-105)(295,-55){4}{5}
\Vertex(225,-105){1.5}
\Vertex(225,-55){1.5}
\Vertex(295,-105){1.5}
\Vertex(295,-55){1.5}
\ArrowLine(150,20)(165,35)  
\Text(90,5)[r]{\scriptsize{$\vec{K}$}}
\ArrowLine(165,35)(180,20)
\Text(110,5)[l]{\scriptsize{$\vec{K}+\vec{Q}$}}
\ArrowLine(150,120)(165,105)
\Text(110,77)[l]{\scriptsize{$\vec{K}'-\vec{Q}$}}
\ArrowLine(165,105)(180,120)
\Text(90,77)[r]{\scriptsize{$\vec{K}'$}} 
\Photon(165,35)(165,48.7){4}{1}
\Photon(165,105)(165,91.15){4}{1}
\Text(110,37)[l]{\scriptsize{$(b)$}}
\Vertex(165,35){1.5}
\Vertex(165,105){1.5}
\CArc(143.75,70)(30,-45,45)
\LongArrowArc(143.75,70)(30,-45,5)
\CArc(186.25,70)(30,135,225)
\LongArrowArc(186.25,70)(30,135,185)
\Vertex(165,92.15){1.5}
\Vertex(165,48.7){1.5}
\end{picture}
\end{center}
\caption{\footnotesize{ First and second order graphs contributing to 
$\Gamma$. We use the notations $\sharp=\{ \; ,'\}$ and 
$\bar \sharp=\{',\, \}$.}}
\label{graph1-vertex} 
\end{figure}
First the Landau function will be evaluated to second order in $U$. 
Corrections from the $z_k$ factors are at least second order in $U$, 
therefore setting
$z_k=1$ is fully consistent up to second order. The same argument justifies
the neglect of effective mass corrections. Finally the second order
quasiparticle interaction function reads
{\setlength \arraycolsep{2pt} 
  \begin{eqnarray}  \label{f_second_order_upup}
    f_{\uparrow \uparrow}(\vec{k}_F,\vec{k}'_F)
    &=&f_{\downarrow \downarrow}(\vec{k}_F,\vec{k}'_F) \nonumber \\
    &=&U^2 \chi_{PH}(\vec{k}_F-\vec{k}'_F), \\
    \nonumber \\
    f_{\uparrow \downarrow}(\vec{k}_F,\vec{k}'_F)
    &=&f_{\downarrow \uparrow}(\vec{k}_F,\vec{k}'_F) \nonumber \\
    &=&U+U^2
(\chi_{PP}(\vec{k}_F+\vec{k}'_F)+\chi_{PH}(\vec{k}_F-\vec{k}'_F)),\nonumber \\
  \end{eqnarray}
with 
{\setlength \arraycolsep{2pt} 
  \begin{eqnarray} 
    \label{chiPH}
    \chi_{PH}(\vec{q}_0)&=& \frac{1}{( 2\pi )^2} \int d^2p \,
\frac{n_{\vec{p}}-n_{\vec{q}_0+\vec{p}}}
    {\epsilon_{\vec{q}_0+\vec{p}}-\epsilon_{\vec{p}}}.
  \end{eqnarray} 
and
{\setlength \arraycolsep{2pt} 
  \begin{eqnarray} 
    \chi_{PP}(\vec{q}_0)&=& \frac{1}{( 2\pi )^2} \int d^2p \,
\frac{1-n_{\vec{p}}-n_{\vec{q}_0-\vec{p}}}{2\mu -
      \epsilon_{\vec{p}}-\epsilon_{\vec{q_0}-\vec{p}}}. \nonumber \\
  \end{eqnarray} 
Here $n_p=1/(1 + \exp(\epsilon_p-\mu)/T)$ is the Fermi
distribution at temperature $T$.

For the analysis of the physical properties of the interacting Fermi liquid 
it is convenient to introduce symmetric ($f^s$) and  antisymmetric parts 
($f^a$) of the Landau quasiparticle interaction by
  \begin{equation}  
    f^{s,a} = \frac{1}{2}(f_{\uparrow \uparrow}\pm f_{\downarrow \uparrow}).
  \end{equation}  

$f^s(\vec{k}_F, \vec{k}'_F)$ and $f^a(\vec{k}_F, \vec{k}'_F)$ are
computed for a fixed value of the chemical potential $\mu$.
$\vec{k}_F$ and $\vec{k}'_F$ are parame\-trized by the two angles $\theta$ and
$\theta'$, measured from the symmetry axis $\Gamma X$, as shown in the inset 
of Fig. \ref{loc_suc_2o}. Fig.\ref{Land_int_grap} shows the results
obtained for the parameters $t'/t = 0.3$, $\mu/t'=-3.83$, 
$T/t=0.002$ and $U/t=1$. Note that for this choice of the chemical potential
$\mu/t'>-4$, the Fermi surface consists of four arcs which lead to the breaks
in the curves. The most prominent feature is the divergence along a diagonal
line with $\pi < \theta' < 3\pi/2$ in both functions. This reflects simply the
cooper divergence in second order perturbation theory and it is removed by
summing the ladder graphs in the particle-particle channel as discussed below.
\begin{figure}[h] 
  \begin{center}
    \includegraphics[width=.5 \textwidth]{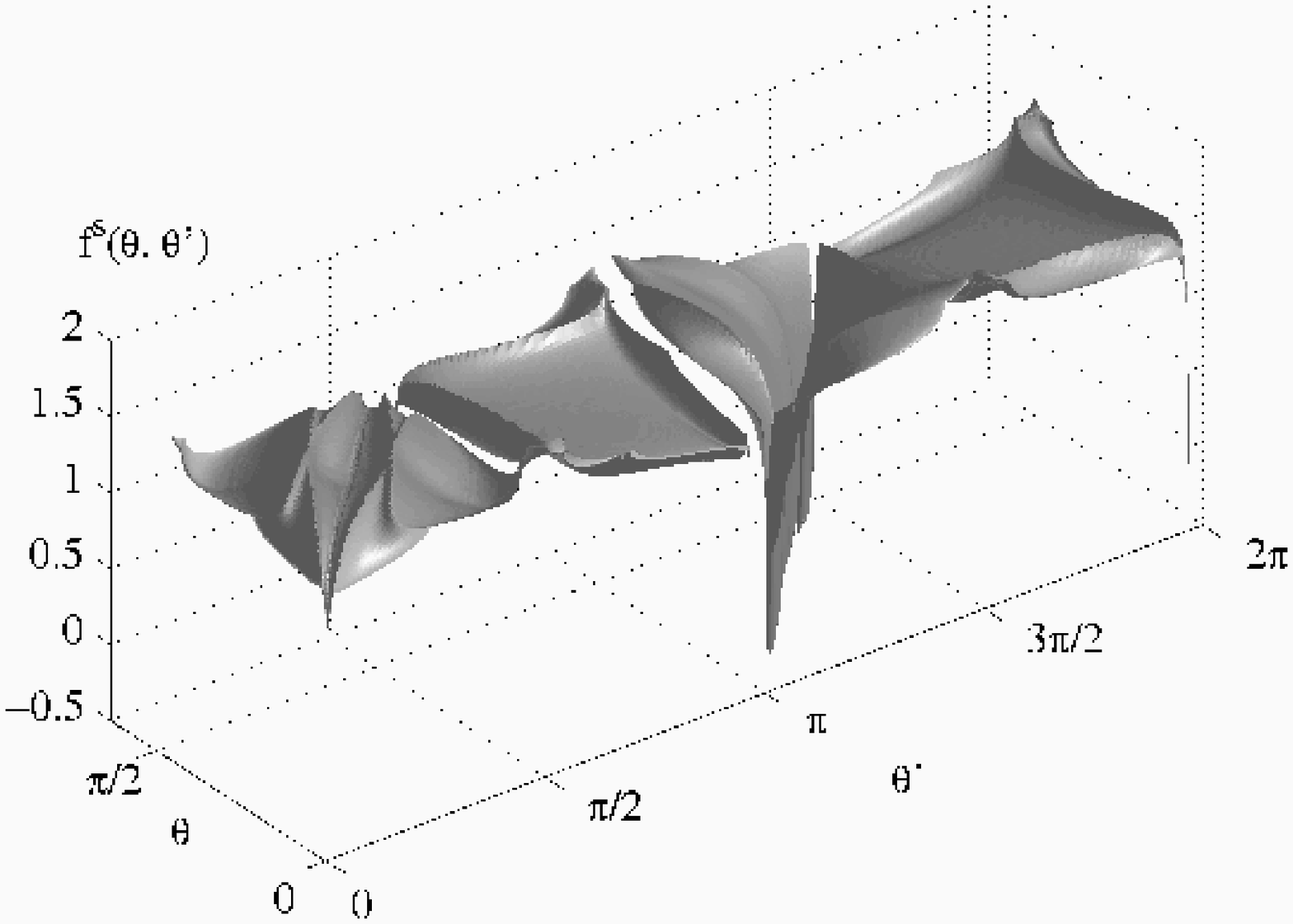} 
    \includegraphics[width=.5 \textwidth]{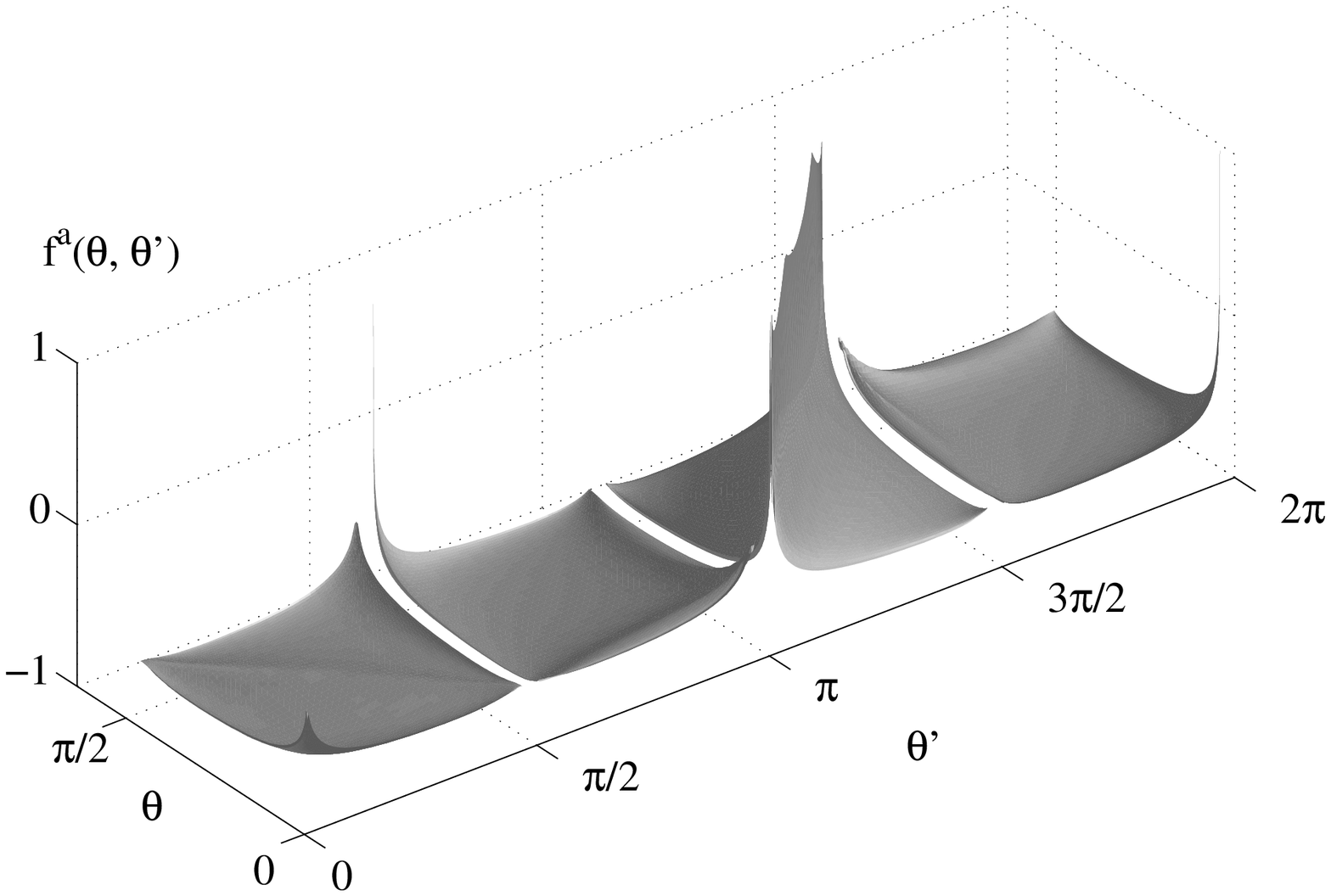} 
    \caption{\footnotesize{ Numerical results for the symmetric 
 $f^s(\theta,\theta')$
    and antisymmetric part  $f^a(\theta,\theta')$  of the Landau
    interaction obtained by the second order perturbation approximation  for
    the parameters $t'=0.3t$, $\mu=-3.83t'$, $U=1t$ and $T=0.002t$. To the second order $f^{a,s}(\vec{k}_F,-\vec{k}_F)$ show unphysical
    divergences (sharp diagonal features for $\pi < \theta' < 3\pi/2$).}}   
  \label{Land_int_grap}
  \end{center}
\end{figure}

\section{Total charge and spin susceptibility for the almost isotropic case}
%
%
In a Fermi liquid the static uniform charge and spin susceptibilities
$X_c$ and $X_s$ involve only excitations in the neighborhood
of the Fermi surface. Therefore they can be calculated as Fermi surface 
integrals over
$\vec{k}$-space local quantities $\chi_c(\vec{k}_F)$ and $\chi_s(\vec{k}_F)$
which define local susceptibilities for Fermi surface points described by the
wavevector $\vec{k_F}$,
  \begin{eqnarray}  
    \rho^2 X_c &=& \frac{1}{\Omega} \frac{dN_e}{d\mu}= 
    \int dl_F 
    \, \chi_c(\vec{k}_F){} \label{chic}, \\
    \frac{X_s}{\mu_B^2} &=& \frac{1}{\mu_B^2 \Omega} \frac{dM}{dH}= \int
    dl_F \, \chi_s(\vec{k}_F).  \label{chis} 
  \end{eqnarray}  
Here, $\int  dl_{F}$ denotes the line
integral along the Fermi surface and $\Omega$ is the total volume. 
$N_e$ denotes the number of electrons and $\rho=N_e/\Omega$ denotes their density. $\mu_B$ is the Bohr magneton.
The $\vec{k}$-space local susceptibilities are solutions of two
inhomogeneous linear equations,

\begin{equation} \label{dkfdmu}  
  L^{s}( \chi_{c})=2, \qquad  L^{a}(\chi_{s})=2,
\end{equation}
which express via the two linear integral operators
$L^{s,a}$,
%
  \begin{eqnarray}
    \frac{L^{s,a}(\chi)}{(2\pi)^2}&=& v_F \chi(\vec{k}_F) \,  
    + \int \frac{dl_F'}{2\pi^2} \; f^{s,a}(\vec{k}_F,\vec{k}'_F)\,
    \chi(\vec{k}'_F), \nonumber  \label{lineq} \\ 
  \end{eqnarray}  
how the local susceptibilities are renormalized by the Landau
$f$-function \cite{theory-of-FL1}.
Here, $v_F \equiv v_{\vec{k}_F}$ denotes the magnitude of the Fermi velocity   
$\vec{v}_{\vec{k}_F}=\partial \epsilon_{\vec{k}}/ \partial \vec{k}|_{\vec{k}=\vec{k}_F}$    
of a quasiparticle with wavevector $\vec{k}$.  
For an isotropic Fermi surface and isotropic interactions, 
$\chi_c(\vec{k})$ and $\chi_s(\vec{k})$ are obviously two constants 
independent of $\vec{k}$. This
allows one to express $X_c$ and  $X_s$ using
the conventional Landau parameters $F_{0}^{s,a}$, 
which are defined as  zeroth order coefficients in the expansion 
of the Landau interaction functions ${f^{s,a}}$ in normalized 
Legendre polynomials. 
In the anisotropic case the Landau interaction functions can 
be expanded in a similar way,
\begin{equation}  
  f^{s,a}(\vec{k}_F,\vec{k}'_F)=\frac{1}{N_F}\sum_{n=0}^{\infty}
  \sum_{n'=0}^{\infty} F_{n,n'}^{s,a} \psi_n(\vec{k}_F)  \psi_{n'}(\vec{k}'_F)
  \, , 
\end{equation}       
where $\psi_0(\vec{k}_F)=1$, $\psi_n$ are orthogonal functions on the Fermi 
surface and $N_F$ is the density of states at the Fermi surface. Note that
 in general
the off-diagonal terms $F_{n,n'}^{s,a}= F_{n',n}^{s,a}$ do not vanish when 
both $\psi_n$ and $\psi_n'$ belong to the same irreducible representation 
of the crystal symmetry group.

Initially, let us assume that the isotropic components of $f^{s,a}$,
\begin{equation}  
  \label{Landau_par}
  F_0^{s,a}=\frac{ \int \int \frac{dl_{F}}{v_F} 
    \frac{dl'_{F}}{v'_F}
    f^{s,a}(\vec{k}_F, \vec{k'}_F)} {2\pi^2 \int 
    \frac{dl_F}{v_F}} \equiv F_{0,0}^{s,a},
\end{equation}    
are dominant. This assumption was also made
in Ref. \cite{fuseya}. In this case the compressibility (\ref{chic}) 
simplifies to
{\setlength \arraycolsep{2pt} 
  \begin{eqnarray}  
    \rho^2 X_c &=& N_F -  F^s_0 
    \int dl'_{F} \chi_c(\vec{k}'_F){} \nonumber \\
    &=& N_F -  F^s_0 \rho^2 X_c{} \, . \nonumber  \\ 
  \end{eqnarray}  
Thus we arrive at the well-known result for an isotropic Fermi liquid,
 \begin{equation}    \rho^2 X_c(\mu) = \frac{ N_F}{1+F^s_0(\mu)} \, . \label{comp_isot} \end{equation}
Using the same simplification we obtain for the spin susceptibility 
{\setlength \arraycolsep{2pt} 
  \begin{eqnarray}  
    \frac{X_s(\mu)}{\mu_B^2} &=& \frac{ N_F }{1+F^a_0(\mu)}.  \label{susc_isot}
  \end{eqnarray}  
Fuseya et al. \cite{fuseya} evaluated charge and spin susceptibilities 
for the Hubbard model with  $t'=0$ assuming that 
the isotropic components of $f^{s,a}$ were dominant.  
Their results showed a reduction of the uniform spin susceptibility 
for $U>2t$ at density near to half-filling. It turns out that this effect 
is tied the van Hove band filling where the density of states at the 
Fermi surface diverges due to van Hove singularities at the 
saddle points $(\pi, 0 )$ and $(0,\pi)$.
In the special case $t'=0$ the reduction of the spin susceptibility
 does not have a simple physical interpretation. 
In this case the van Hove filling coincides 
with half-filling, where the system is believed to be an antiferromagnetic 
insulator and the Fermi liquid approach must break down at low energy scales. 
But for finite next-nearest neighbor hopping $t'\not= 0$ 
the van Hove filling does not coincide anymore with
the half-filled or perfectly nested case and in the vicinity of this
filling the system may be in a paramagnetic state. In this case a
reduction of the uniform spin susceptibility 
for $U>2t$ could be interpreted as an indication for an opening of a
spin gap in the Hubbard model close to half filling. 
\begin{figure}[h] 
  \begin{center}
    \includegraphics[width=.5 \textwidth]{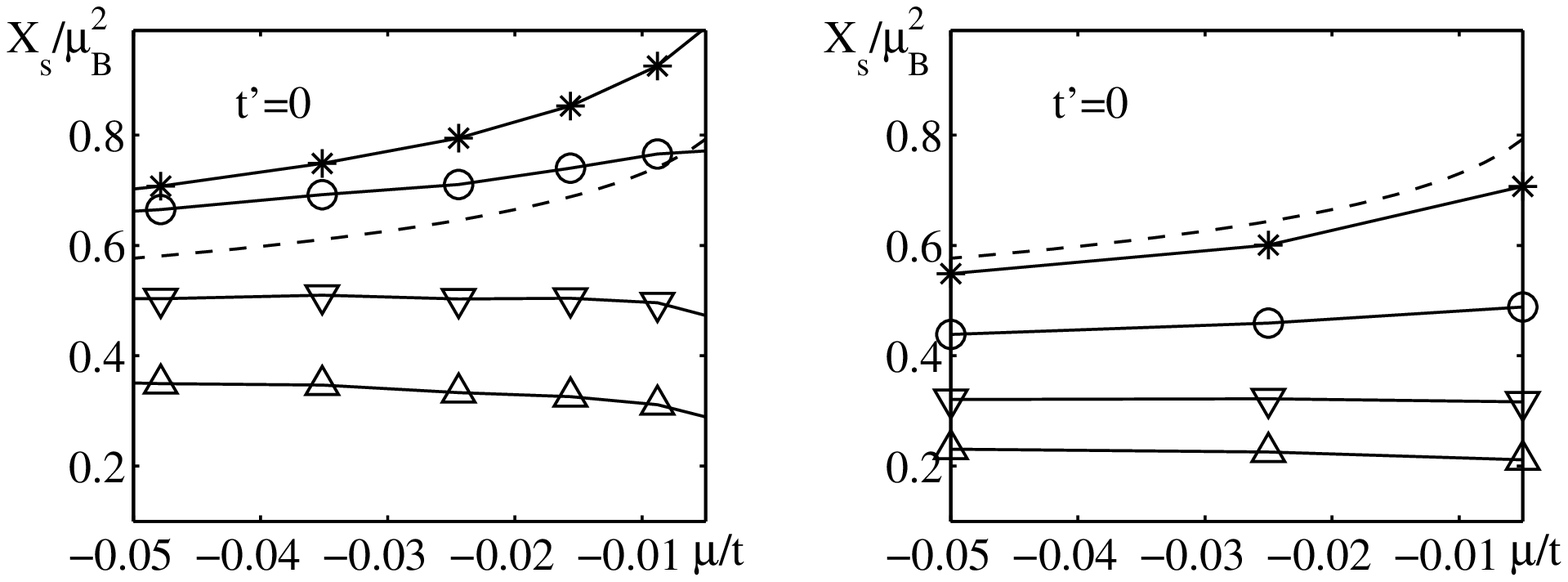} 
    \includegraphics[width=.5 \textwidth]{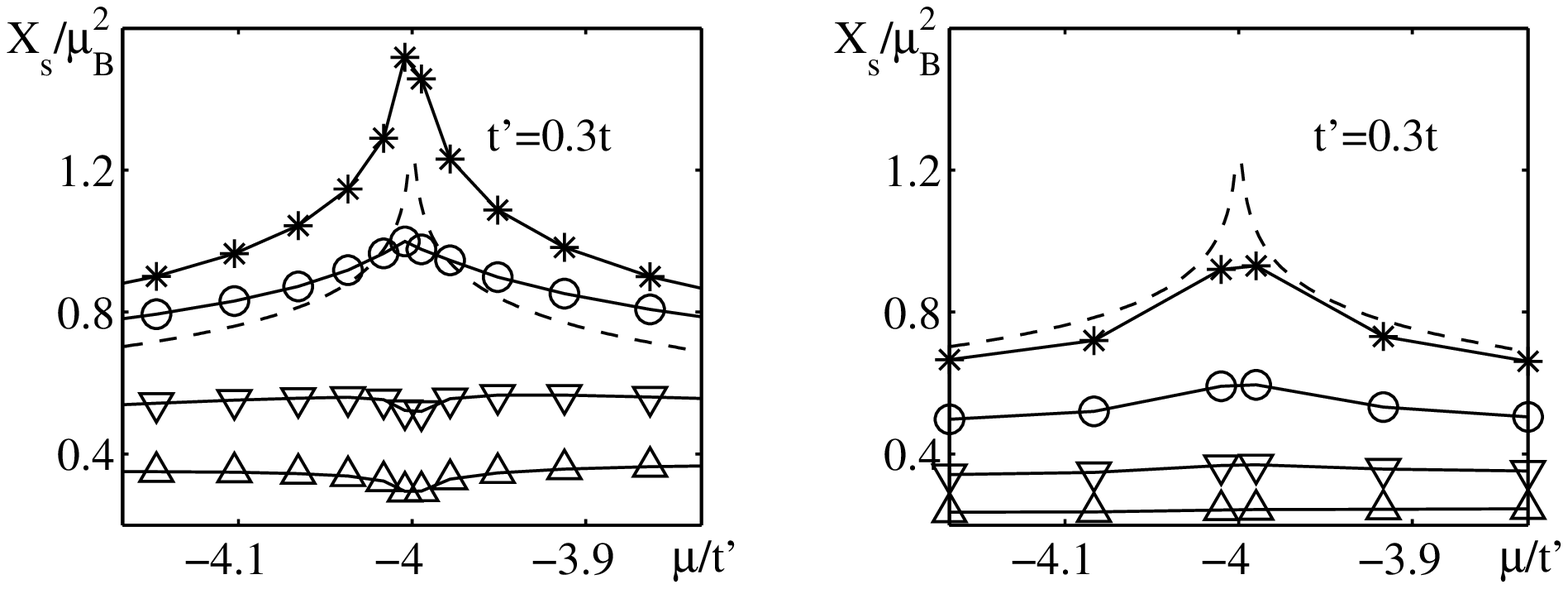} 
    \caption{\footnotesize{Comparison between the uniform spin 
        susceptibility $X_s$ as function of the chemical potential $\mu$ 
        obtained from the approximate relations (\ref{susc_isot}) 
        (plots on the left) and 
       from the solution of the full integral equations 
        (\ref{dkfdmu},\ref{chis}) (plots on the right),
        for different values of interaction $U$: $*$ $U/t=1$, $\circ$ $U/t=2$, 
        $\bigtriangledown$ $U/t=3$, $\bigtriangleup$
        $U/t=4$. For all plots $T/t=0.002$. The susceptibility for the 
non-interacting system ($U/t=0$) is shown as dashed line. 
        }}
    \label{comparison}
\end{center}
\end{figure}   
However, our results obtained by 
solving the full integral equations (\ref{dkfdmu}), without the
assumption  of a dominance of the $n=0$ terms,  unfortunately remove this
promising finding. 
This is shown in Fig. \ref{comparison}, where the data assuming that 
the isotropic terms are dominant (left panels) are compared 
with those obtained by
solving (\ref{chis}, \ref{dkfdmu}) (right panels). In particular in 
the two figures to the right 
the spin susceptibility does not decrease in any significant way as the van
Hove filling ($\mu=0$ for $t'=0$ and $\mu/t'=-4$ for $t'=0.25t$) is 
approached. For the charge susceptibility the qualitative behavior of the
results obtained by solving (\ref{chic}, \ref{dkfdmu}) is compatible 
with the one obtained by Fuseya, even if a quantitative difference is 
present, see Fig.
\ref{comparison1}.
\begin{figure}[h] 
  \begin{center}
    \includegraphics[width=.5 \textwidth]{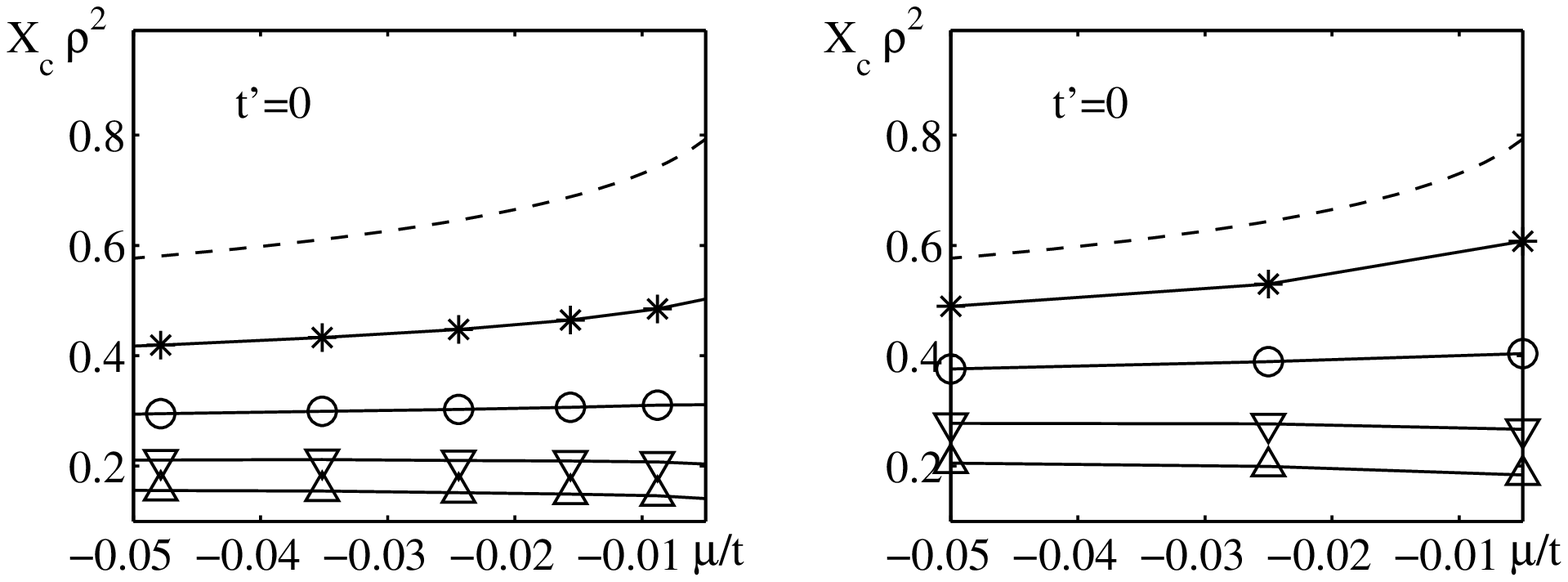} 
    \includegraphics[width=.5 \textwidth]{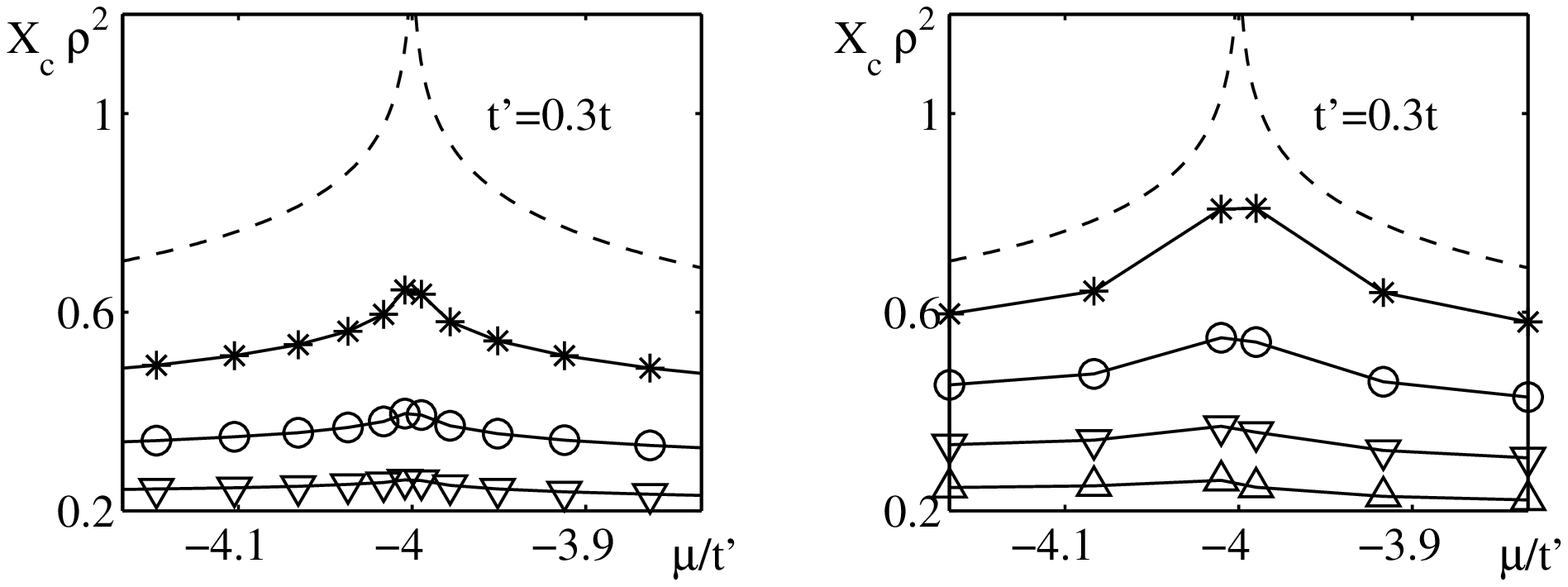} 
    \caption{\footnotesize{Same as in Fig. \ref{comparison}, but for
        the charge susceptibility $X_c$.
        }}
    \label{comparison1}
\end{center}
\end{figure}   

Thus the conclusion of Fuseya et al.\cite{fuseya} that indications
for a spin gap in the 2D Hubbard model can already be observed in second
order perturbation theory seems to be an artifact of their
approximation. The solution of the full integral equation reveals that 
the assumption of a dominant $l=0$ Landau parameter is not 
always justified.
%
\section{Local spin and charge susceptibilities in $\vec{k}$-space}
%
Since the Landau interaction functions are very anisotropic,
it is interesting to analyze how the local values of the static
susceptibilities change with the interaction $U$. As mentioned in the introduction, we are particularly interested in the question whether 
there are tendencies 
towards a Fermi surface truncation, i.e. the formation of incompressible 
$\vec{k}$-space regions at the saddle points, as suggested by one-loop 
renormalization group studies, already within a low order Fermi liquid treatment.
For the non-interacting system, 
$\chi_c(\vec{k}_F)$ and $\chi_s(\vec{k}_F)$ are proportional 
to  the inverse velocity $1/v_{\vec{k}_F}$.
Since $v_{\vec{k}_F}$ vanishes at the saddle points, 
a divergence appears in the local spin and charge 
susceptibilities around these points at the
van Hove filling - a behavior opposite to incompressibility.
The  results for the interacting system obtained with the second order 
Landau function show a general 
reduction of the local values of the susceptibilities with increasing 
interaction strength $U$.
\begin{figure}[h] 
  \begin{center}
    \includegraphics[width=.5 \textwidth]{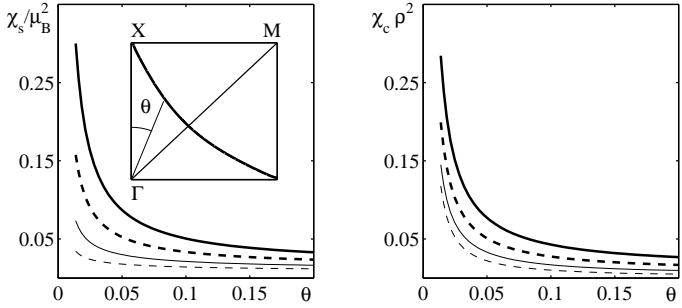}
    \caption{\footnotesize{$\vec{k}$-space local values of $\chi_s(\vec{k}_F)$
        and $\chi_c(\vec{k}_F)$ for $\vec{k}_F$ near a saddle point
        for different values of  $U$: $U/t=1$ thick line, $U/t=2$  thick dashed
        line, $U/t=3$ thin line, $U/t=4$ thin dashed line. The inset
        shows how the angle $\theta$ is used to parameterize the points 
        of the Fermi surface ($\vec{k}_F(\theta)$). Results computed for
        $t'=0.3$, $\mu/t'=-3.99$, and $T/t=0.002$.}}
    \label{loc_suc_2o}
\end{center}
\end{figure}        
Fig. \ref{loc_suc_2o} shows $\chi_s(\vec{k}_F)$ and $\chi_c(\vec{k}_F)$ 
when $\vec{k}_F$ is close to the saddle point. The inset in the spin
susceptibility graph indicates how the angle $\theta$ is used to 
parameterize the points of the Fermi surface, $\vec{k}_F(\theta)$.
Particularly interesting is the fact that near the saddle point the reduction
of the local spin susceptibility is pronounced. 
The charge susceptibility shows a similar behavior, but in a less marked way.
\begin{figure}[h] 
  \begin{center}
    \includegraphics[width=.48 \textwidth]{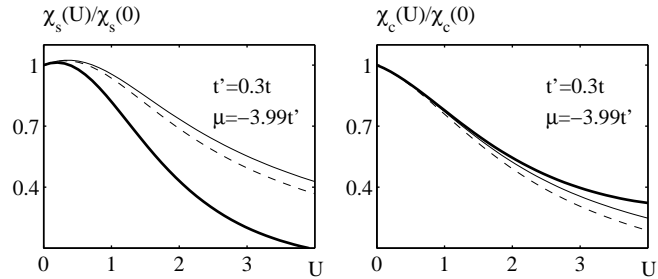} 
    \caption{\footnotesize{Relative variations of the $\vec{k}$-space 
local spin and charge susceptibilities ($\chi_s(U)$ and  $\chi_c(U)$) near the van Hove 
        filling for three different values of the angle $\theta$ as 
        function of the interaction $U$. $\theta=\theta_{min}$ thick line,
        $\theta \approx \pi/8$ thin dashed line, $\theta=\pi/4$
        thin line. T/t=0.002. }}
    \label{susc_rel197}
\end{center}
\end{figure}   
In fact the comparison between
the two graphs of the Fig. \ref{susc_rel197} shows that the relative value of
$\chi_s(\vec{k}_F)$ decreases more rapidly near the saddle points
($\theta=\theta_{min}$) than at other points on the
Fermi surface (e.g. $\theta \approx \pi/8$ and $\theta=\pi/4$), while for $\chi_c(\vec{k}_F)$ the situation is reversed. 
\begin{figure}[h] 
  \begin{center}
    \includegraphics[width=.48 \textwidth]{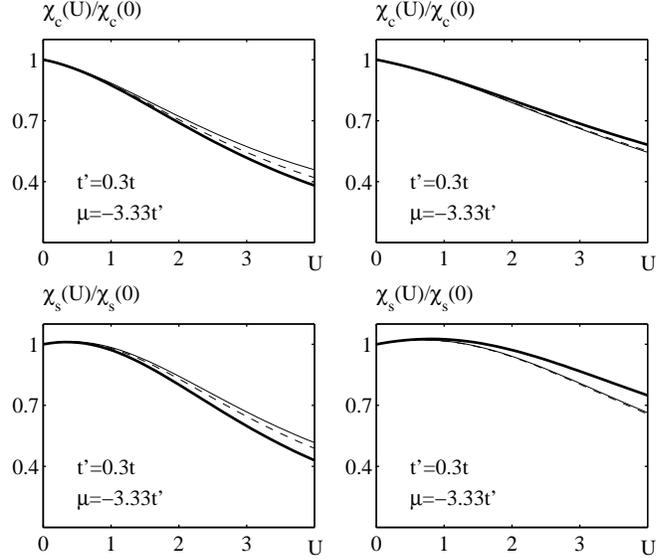}  
    \caption{\footnotesize{Left plots: relative 
        variation of the local spin and charge  
        susceptibilities ($\chi_s(U)$,  $\chi_c(U)$) for three different 
        values of the angle $\theta$ as 
        function of the interaction $U$. $\theta=\theta_{min}$ thick line,
        $\theta \approx \pi/8$ thin dashed line, $\theta=\pi/4$
        thin line. $T/t=0.002$. The band filling is larger than the van Hove filling and elastic umklapp processes at the Fermi surface are possible.
 Right plots: results obtained excluding 
all umklapp processes from the Landau interaction
        functions. }}
    \label{susc_rel}
  \end{center}
\end{figure}    

The behavior of the susceptibilities for values of $\mu$ further away from the van Hove filling is noteworthy, too. 
In particular the qualitative difference 
of the susceptibilities between the case $\mu > -4t'$ (density larger than van Hove filling) and $\mu < -4t'$ (density smaller than van Hove filling), deserves to be noticed.
In the first case the two graphs on the left in Fig. \ref{susc_rel} show
that the relative reduction of the local susceptibilities 
is stronger for Fermi surface points near to $X M$ ($\theta=\theta_{min}$), compared to Fermi surface points closer to the diagonal $\Gamma M$ ($\theta \approx \pi/8$, 
$\theta=\pi/4$). The situation
changes if the calculations are done excluding umklapp processes of all kinds 
from the second order contributions, see the two graphs on the right 
in Fig. \ref{susc_rel}. This last fact already indicates the importance of elastic umklapp scattering at the Fermi surface that is allowed for densities larger than the van Hove density.
\begin{figure}[h] 
  \begin{center}
    \includegraphics[width=.48 \textwidth]{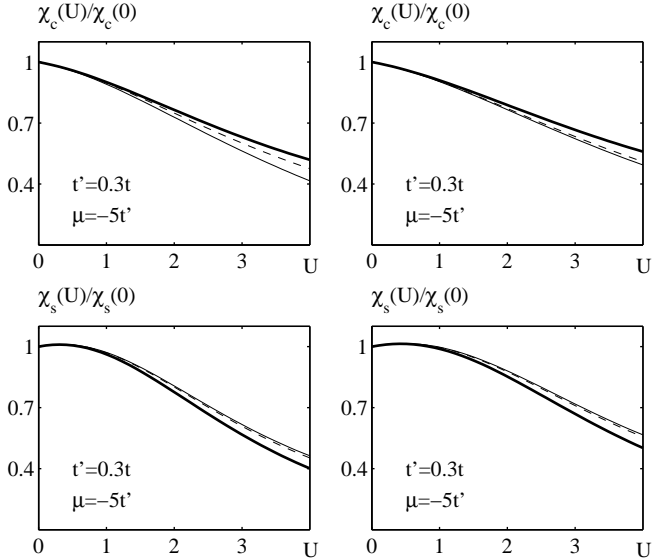} 
    \caption{\footnotesize{Same as in Fig. \ref{susc_rel} for band filling less than the van Hove filling such that elastic umklapp processes at the Fermi surface are absent. }}
 \label{susc_rel1}
 \end{center}
\end{figure}
The second case is when $\mu < -4t'$ 
and elastic umklapp scattering does not enter the low energy physics. 
The two graphs on the left in
Fig. \ref{susc_rel1} show a qualitative behavior of the local spin
susceptibility that is similar to that observed for $\mu > -4t'$. 
But the strongest reduction of the local charge susceptibility is now shifted
to $\theta \approx \pi/8$ and $\theta=\pi/4$ and the role of the umklapp processes becomes marginal at these densities.

%
\section{ Ladder summations}
%
The results in the previous sections were 
obtained with the Landau interaction function
calculated to second order in $U$. Although we observed a general 
suppression of the charge and spin susceptibility with respect to 
the non-interacting values, the effects were weak and no clear 
tendencies towards partial incompressibility could be found. 
In this section we extend the perturbative calculation of the 
Landau function to summations of infinite numbers of diagrams. 
Our motivation for doing this is twofold: {\em a)}
A severe defect of the second order approach is that
the contribution of the $s$-wave pairing channel to the interaction
between the quasiparticles is typically overestimated to this order. 
In fact $f_{a}(\vec{k}_F,\vec{k}'_F)= -U^2\chi_{PP}
(\vec{k}_F+\vec{k}'_F)$ diverges at $T=0$ when $\vec{k}'_F 
\rightarrow -\vec{k}_F$ due to the singular behavior
of the Cooper graph in systems with parity $\epsilon(\vec{k}) =
\epsilon(-\vec{k})$, that is clearly visible from the Fig. 
\ref{Land_int_grap}. In the following we will remedy this defect by summing up the ladder diagrams generated by the particle-particle 
graphs, the result is shown in the Fig. \ref{LAD_int_grap}. 
In this way for $U>0$ the net contribution of the Cooper channel 
at $T=0$ will be zero.
\begin{figure}[h] 
  \begin{center}
    \includegraphics[width=.499 \textwidth]{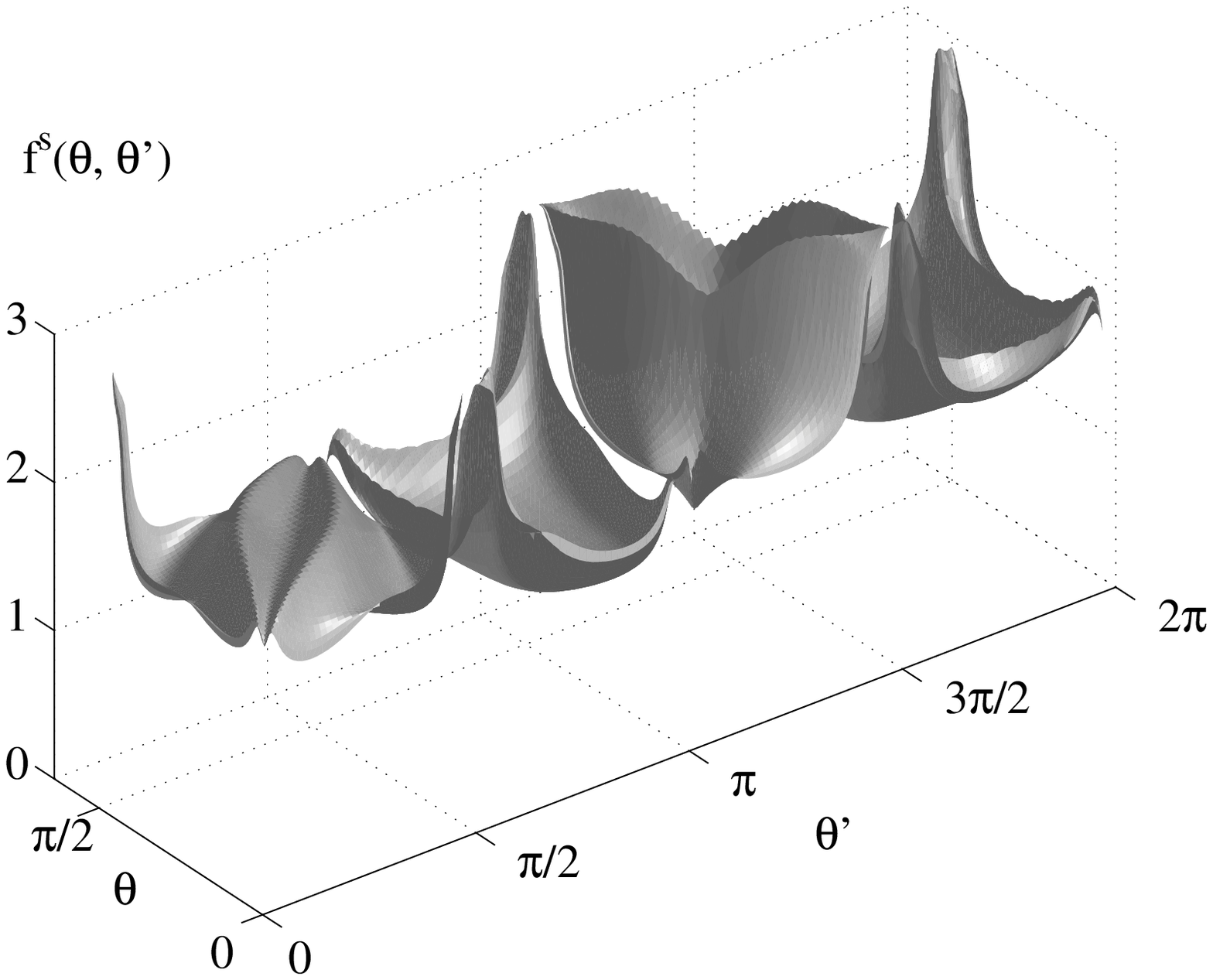}  
    \includegraphics[width=.5 \textwidth]{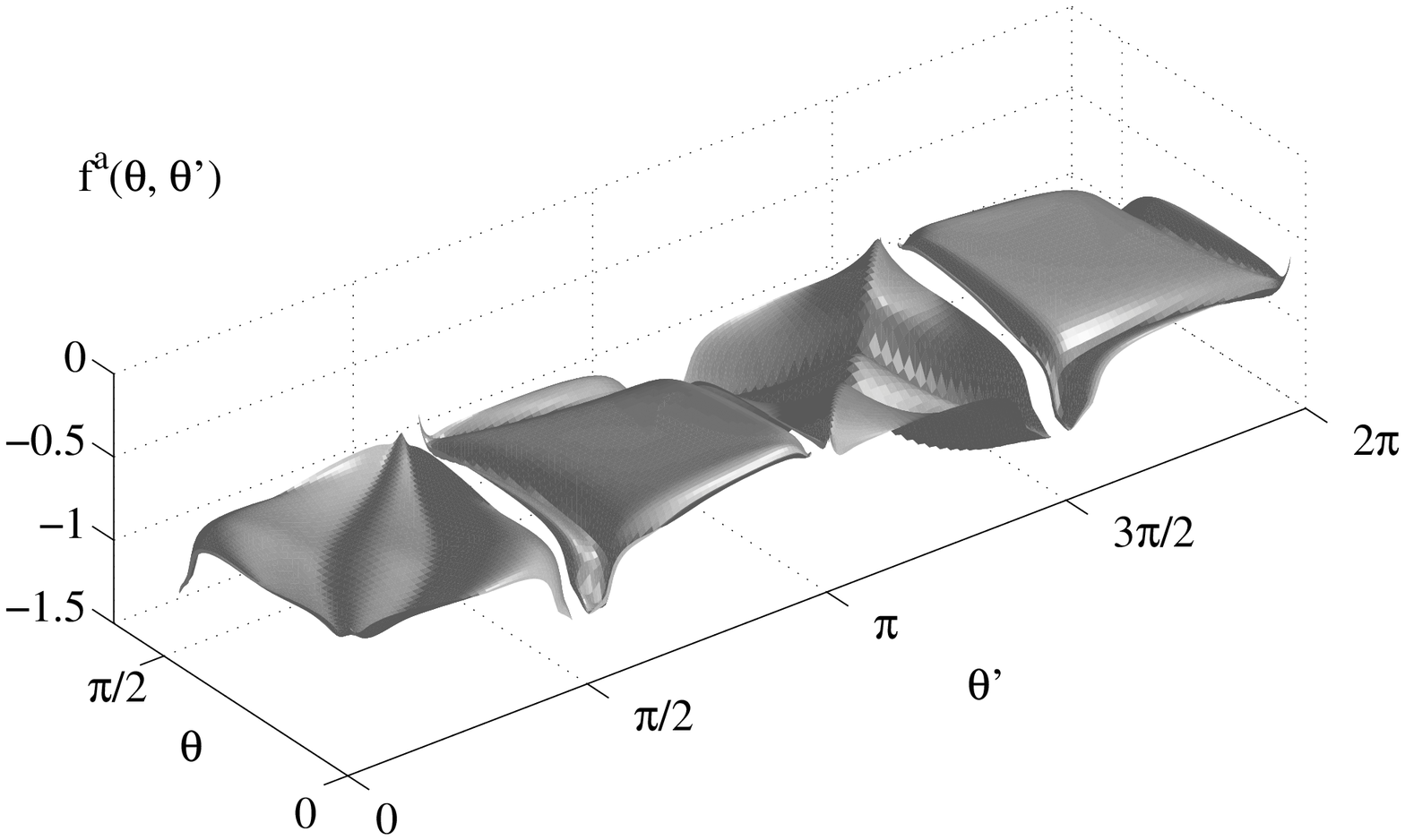}  
    \caption{\footnotesize{Numerical solution of the 
        symmetric $f^s(\theta,\theta')$ and antisymmetric 
        part  $f^a(\theta,\theta')$ of the Landau
        interaction obtained by the ladder and bubble summation for 
        $t'=0.3t$, $\mu=-3.83t'$,  $U=1.5t$ and $T=0.04t$. 
The unphysical
    divergences of $f^{a,s}(\vec{k}_F,-\vec{k}_F)$ found with the second order
    approximation in Fig. \ref{Land_int_grap} are suppressed by the ladder summation.}}   
    \label{LAD_int_grap}
  \end{center}
\end{figure}
{\em b)}
Furthermore we know that the Hubbard model close to half filling is close to
an antiferromagnetic instability. This effect can be modeled in a perturbative
way by summing up an infinite number of bubble and ladder diagrams in the
particle-hole channel, as described below. By this we can enhance the
proximity to the AF instability and simultaneously study whether the local charge and spin susceptibilities exhibit significant changes. 
We should add that our calculation does not aim at a quantitative determination of the phase diagram of the model. Therefore we do not worry too much about the question whether we neglect certain contributions by selecting this particular class of diagrams. Our goal is rather to include and enhance a specific type of scattering processes and to determine the effects of these processes on the Fermi liquid interaction and derived quantities. In the same spirit we neglect possible self energy corrections in these calculations. Within the Fermi liquid framework this concerns mainly the effective mass of the quasiparticles, but a consistent treatment should also take into account  the Fermi surface shift caused by the interactions and lifetime effects on the virtual excitations in the ladder diagrams. However, since generally in spatial dimensions higher than one the self energy does not develop singularities without clear signs in the interactions and since we seek for qualitative results that do not depend on details of the model, we ignore the selfenergy effects and focus on the interaction function. This has the advantage that the analysis remains relatively simple and lucid. 

The graphs contributing to the particle-particle 
and particle-hole ladder and bubble summations are shown
in Fig. \ref{graph2-Ladder}. The only graph
contributing to $f_{\uparrow,\uparrow}(\vec{k}_F,\vec{k}'_F)$ to the second
order is the one-loop particle-hole bubble. This implies that the 
ladder summation for
$f_{\uparrow,\uparrow}$ involves only odd powers of the bubble graph.    
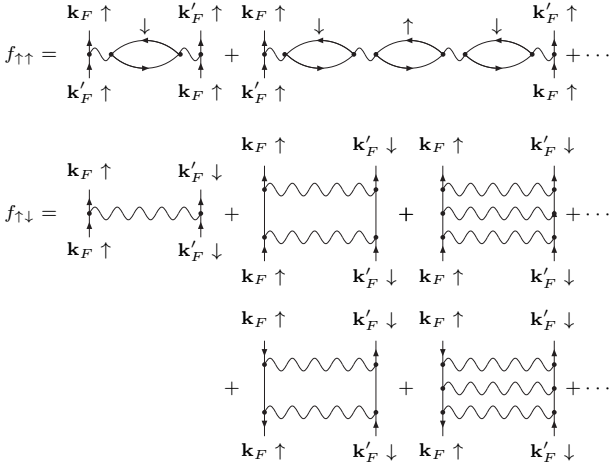
\begin{figure}[h] 
\begin{center} \begin{picture}(370,200)(-20,-60)
\SetScale{0.6}
\ArrowLine(35,150)(35,165)  
\Text(9,99)[r]{\scriptsize{$f_{\uparrow \uparrow}=$}}
\Text(21,115)[c]{\scriptsize{$\vec{k}_F \uparrow$}}
\ArrowLine(35,165)(35,180)
\Text(21,85)[c]{\scriptsize{$\vec{k}'_F \uparrow$}}
\ArrowLine(105,150)(105,165)
\Text(63,115)[c]{\scriptsize{$\vec{k}'_F \uparrow$}}
\ArrowLine(105,165)(105,180)
\Text(63,85)[c]{\scriptsize{$\vec{k}_F \uparrow$}} 
\Photon(35,165)(48.7,165){4}{1}
\Photon(105,165)(91.15,165){4}{1}
\Text(42,110)[c]{\scriptsize{$\downarrow$}}
\Vertex(35,165){1.5}
\Vertex(105,165){1.5}
\CArc(70,143.75)(30,45,135)
\LongArrowArc(70,143.75)(30,45,95)
\CArc(70,186.25)(30,225,315)
\LongArrowArc(70,186.25)(30,225,275)
\Vertex(92.15,165){1.5}
\Vertex(48.7,165){1.5}
\Text(75,99)[c]{\scriptsize{$+$}}
%
\ArrowLine(145,150)(145,165)
\Text(87,115)[c]{\scriptsize{$\vec{k}_F \uparrow$}}
\ArrowLine(145,165)(145,180)
\Text(87,85)[c]{\scriptsize{$\vec{k}'_F \uparrow$}}
\Text(197.2,115)[c]{\scriptsize{$\vec{k}'_F \uparrow$}}
\Text(197.2,85)[c]{\scriptsize{$\vec{k}_F \uparrow$}} 
\Photon(145,165)(158.7,165){4}{1}
\Photon(215,165)(201.15,165){4}{1}
\Text(108,110)[c]{\scriptsize{$\downarrow$}}
\Text(141.6,110)[c]{\scriptsize{$\uparrow$}}
\Text(175.2,110)[c]{\scriptsize{$\downarrow$}}
\Vertex(145,165){1.5}
\Vertex(215,165){1.5}
\CArc(180,143.75)(30,45,135)
\LongArrowArc(180,143.75)(30,45,95)
\CArc(180,186.25)(30,225,315)
\LongArrowArc(180,186.25)(30,225,275)
\Vertex(201.15,165){1.5}
\Vertex(157.7,165){1.5}
\Photon(271,165)(257.15,165){4}{1}
\Vertex(271,165){1.5}
\CArc(236,143.75)(30,45,135)
\LongArrowArc(236,143.75)(30,45,95)
\CArc(236,186.25)(30,225,315)
\LongArrowArc(236,186.25)(30,225,275)
\Vertex(257.15,165){1.5}
\Photon(327,165)(313.15,165){4}{1}
\Vertex(327,165){1.5}
\CArc(292,143.75)(30,45,135)
\LongArrowArc(292,143.75)(30,45,95)
\CArc(292,186.25)(30,225,315)
\LongArrowArc(292,186.25)(30,225,275)
\Vertex(313.15,165){1.5}
\ArrowLine(327,165)(327,180)
\ArrowLine(327,150)(327,165)
\Text(201,99)[l]{\scriptsize{$+\cdots$}}
\ArrowLine(35,50)(35,65)  
\Text(9,39)[r]{\scriptsize{$f_{\uparrow \downarrow}=$}}
\Text(21,55)[c]{\scriptsize{$\vec{k}_F \uparrow$}}
\ArrowLine(35,65)(35,80)
\Text(21,25)[c]{\scriptsize{$\vec{k}_F \uparrow$}}
\ArrowLine(105,50)(105,65)
\Text(63,55)[c]{\scriptsize{$\vec{k}'_F \downarrow$}}
\ArrowLine(105,65)(105,80)
\Text(63,25)[c]{\scriptsize{$\vec{k}'_F \downarrow$}} 
\Photon(35,65)(105,65){4}{5}
\Vertex(35,65){1.5}
\Vertex(105,65){1.5}
\Text(75,39)[c]{\scriptsize{$+$}}
\Line(145,50)(145,80)
\ArrowLine(145,80)(145,95)
\ArrowLine(145,35)(145,50)
\Text(87,65)[c]{\scriptsize{$\vec{k}_F \uparrow$}}
\Text(87,15)[c]{\scriptsize{$\vec{k}_F \uparrow$}}
\Line(215,50)(215,80)
\ArrowLine(215,80)(215,95)
\ArrowLine(215,35)(215,50)
\Text(129,65)[c]{\scriptsize{$\vec{k}'_F \downarrow$}}
\Text(129,15)[c]{\scriptsize{$\vec{k}'_F \downarrow$}} 
\Photon(145,50)(215,50){4}{5}
\Photon(145,80)(215,80){4}{5}
\Vertex(145,50){1.5}
\Vertex(215,50){1.5}
\Vertex(145,80){1.5}
\Vertex(215,80){1.5}
\Text(141,39)[c]{\scriptsize{$+$}}
\Line(257,50)(257,80)
\ArrowLine(257,80)(257,95)
\ArrowLine(257,35)(257,50)
\Text(154.2,65)[c]{\scriptsize{$\vec{k}_F \uparrow$}}
\Text(154.2,15)[c]{\scriptsize{$\vec{k}_F \uparrow$}}
\ArrowLine(327,50)(327,80)
\ArrowLine(327,80)(327,95)
\ArrowLine(327,35)(327,50)
\Text(196.2,65)[c]{\scriptsize{$\vec{k}'_F \downarrow$}}
\Text(196.2,15)[c]{\scriptsize{$\vec{k}'_F \downarrow$}} 
\Photon(257,50)(327,50){4}{5}
\Photon(257,65)(327,65){4}{5}
\Photon(257,80)(327,80){4}{5}
\Vertex(257,65){1.5}
\Vertex(327,65){1.5}
\Vertex(257,50){1.5}
\Vertex(327,50){1.5}
\Vertex(257,80){1.5}
\Vertex(327,80){1.5}
\Text(141,39)[c]{\scriptsize{$+$}}
\Text(201,39)[l]{\scriptsize{$+\cdots$}}
\SetScaledOffset(0,-110)
\Line(145,50)(145,80)
\ArrowLine(145,95)(145,80)
\ArrowLine(145,50)(145,35)
\Line(215,50)(215,80)
\ArrowLine(215,80)(215,95)
\ArrowLine(215,35)(215,50)
\Photon(145,50)(215,50){4}{5}
\Photon(145,80)(215,80){4}{5}
\Vertex(145,50){1.5}
\Vertex(215,50){1.5}
\Vertex(145,80){1.5}
\Vertex(215,80){1.5}

\Line(327,50)(327,80)
\Line(257,50)(257,80)
\ArrowLine(257,95)(257,80)
\ArrowLine(257,50)(257,35)
\ArrowLine(327,80)(327,95)
\ArrowLine(327,35)(327,50)
\Photon(257,50)(327,50){4}{5}
\Photon(257,65)(327,65){4}{5}
\Photon(257,80)(327,80){4}{5}
\Vertex(257,65){1.5}
\Vertex(327,65){1.5}
\Vertex(257,50){1.5}
\Vertex(327,50){1.5}
\Vertex(257,80){1.5}
\Vertex(327,80){1.5}
\SetOffset(0,-66)
\Text(75,39)[c]{\scriptsize{$+$}}
\Text(87,65)[c]{\scriptsize{$\vec{k}_F \uparrow$}}
\Text(87,15)[c]{\scriptsize{$\vec{k}_F \uparrow$}}
\Text(129,65)[c]{\scriptsize{$\vec{k}'_F \downarrow$}}
\Text(129,15)[c]{\scriptsize{$\vec{k}'_F \downarrow$}} 
\Text(154.2,65)[c]{\scriptsize{$\vec{k}_F \uparrow$}}
\Text(154.2,15)[c]{\scriptsize{$\vec{k}_F \uparrow$}}
\Text(196.2,65)[c]{\scriptsize{$\vec{k}'_F \downarrow$}}
\Text(196.2,15)[c]{\scriptsize{$\vec{k}'_F \downarrow$}} 
\Text(141,39)[c]{\scriptsize{$+$}}
\Text(201,39)[l]{\scriptsize{$+\cdots$}}
\end{picture}
\end{center}
\caption{\footnotesize{Graphs contributing to the Landau interaction
 functions generated by the ladder summations of the first 
 and second order graphs.}}
\label{graph2-Ladder} 
\end{figure}

Thus the Landau interaction functions can be expressed as 
{\setlength \arraycolsep{2pt} 
  \begin{eqnarray}  \label{f_ladd_sum_uu}
    f_{\uparrow,\uparrow}
    &=&\frac{U^2 \chi_{PH}(\vec{k}_F-\vec{k}'_F)}{1-U^2 \chi_{PH}^2(\vec{k}_F-\vec{k}'_F)}, \\
    \nonumber \\
    f_{\uparrow,\downarrow}
    &=&\frac{U}{1-U\chi_{PP}(\vec{k}_F+\vec{k}'_F)}+\frac{U^2 \chi_{PH}
      (\vec{k}_F-\vec{k}'_F)}{1-U\chi_{PH}
      (\vec{k}_F-\vec{k}'_F)}.\nonumber \\
  \label{f_ladd_sum_ud}
  \end{eqnarray}
With the Landau interaction function obtained above Eqns. 
(\ref{dkfdmu}) and (\ref{lineq}) are solved again. Contrary to
the second order approximation the local values of the spin 
susceptibility diverge to $\infty$ for $U>U_c$ with a 
critical $U_c(\mu)$ indicating a
ferromagnetic instability of the system near the van Hove 
filling. This increased tendency towards ferromagnetism 
can be explained by the suppression of singlet pairing   
by the ladder summation. In fact the same unphysical divergence 
of the second order particle-particle diagram was the main 
reason for the suppression of the spin susceptibility in 
the second order results. However, a more sophisticated ladder 
summation using a renormalized pair scattering\cite{scalapino} or an 
renormalization group approach \cite{honerkamp,halboth} will tend to suppress the spin susceptibility again. This is because singlet pairing tendencies 
will be generated in higher angular momentum channels, especially in 
the $d_{x^2-y^2}$-channel.  

The behavior of the local charge susceptibility when the ladder summation is
included is a consistent 
extension of the second order calculation with a stronger 
suppression of $\chi_c(\vec{k}_F)$ 
near the saddle points. The graphs on the left in Fig. \ref{loc_comp_lad} 
show $\chi_c(\vec{k}_F)$ with $\vec{k}_F$ close to a
saddle point for four different values of $U$. At $U=1.62t$ 
( thin dashed line), the local charge susceptibility goes to zero. 
This does not occur if the umklapp processes are excluded from the perturbation expansion, as shown
in the right panel of Fig. \ref{loc_comp_lad}. The comparison
between these two last graphs confirms the fundamental role 
of umklapp processes for the suppression of the charge susceptibility 
near the saddle points. In that sense the ladder summation results 
for the Landau function lead to a similar behavior of the charge 
compressibility as the one observed in one-loop renormalization group treatments.         
\begin{figure}[h] 
  \begin{center}
    \includegraphics[width=.5 \textwidth]{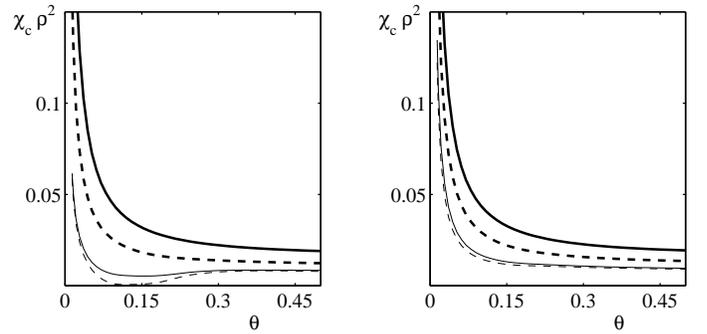} 
    \caption{\footnotesize{Left: $\vec{k}$-space local 
        values of $\chi_c(\vec{k}_F)$ for $\vec{k}_F$ near a saddle 
        point for different values of  $U$: $U/t=0.5$ thick line, 
        $U/t=1.2$  thick dashed line, $U/t=1.91$ thin line, 
        $U/t=2.05$ thin dashed line. Right: results 
        obtained excluding umklapp processes 
from the perturbation expansion. For all data 
        $t'=0.3$, $\mu=-3.99t'$ and $T/t=0.04$. }}
    \label{loc_comp_lad}
\end{center}
\end{figure}

Another similarity with the renormalization group calculations are the tendencies towards
deformations of the Fermi surface, also known as Pomeranchuk instabilities and
originally suggested for the $t$-$t'$ Hubbard model by Halboth and
Metzner\cite{halboth}. Such an instability spontaneously breaks the fourfold
symmetry of the electronic dispersion. Close to the van Hove filling the 
main energy gain comes from lowering the kinetic energy by pushing one saddle point deeper below the Fermi level.
The Landau energy functional for this process can be written, as 
quadratic form in the Fermi surface shifts $\delta s(\vec{k}_F)$,
\begin{equation} \delta E =\frac{1}{(2\pi)^4} \int dl_F L^s(\delta s) \delta s(\vec{k}_F) \, \label{deformationenergy} \end{equation} 
where $L^s$ is defined by (\ref{lineq}).
Negative eigenvalues of the linear integral operator
$L^s$ imply a lowering of the energy (\ref{deformationenergy}) by a suitable deformation of the Fermi surface.
For the specific case $t'/t=0.3$, $\mu/t'=-3.99$ and $T/t=0.04$ 
a negative eigenvalue appears
already at $U/t=1.91$, slightly before 
$\chi_c(\vec{k}_F)$ is suppressed down to zero close to the saddle points at $U/t=2.05$. 
The corresponding eigenvector is
shown in the graph to the left of Fig. \ref{pomeranchuk} and signals a
deformation $\delta s(\theta)$ of the Fermi surface breaking the point
group symmetry of the square lattice, as shown schematically in the plot on the right of the same figure. No attempt was made to estimate the actual magnitude of the Fermi surface deformation.

\begin{figure}[h] 
  \begin{center}
    \includegraphics[width=.5 \textwidth]{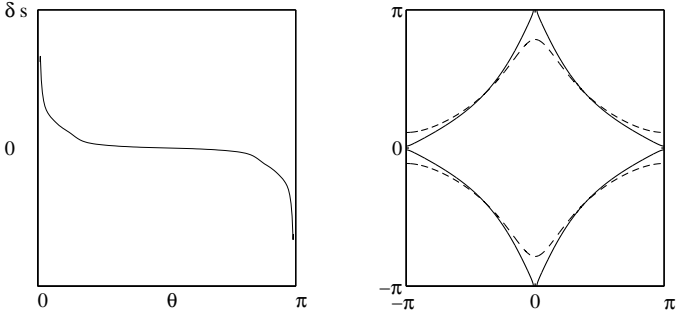} 
    \caption{\footnotesize{Left: unstable eigenvector $\delta
        s(\theta)$ of the operator $T_{\Xi_c^{-1}}$ with 
        eigenvalue $\lambda=-0.01t < 0$ obtained for $t'=0.3t$, $\mu=-3.99t'$, and $U/t=1.91$. Right: 
        Pomeranchuk deformation of the Fermi 
        surface $\propto \vec{v} (\theta) \delta s(\theta)$.}}
    \label{pomeranchuk}
\end{center}
\end{figure}

In the case $t'= 0.3t$, the above scenario is observed for a very large range
of chemical potential values, i.e. between $\mu / t' = -5.66$ and $\mu /t' =-3.33$. Note however that this instability is only one example out of many potential instabilities of the 2D Hubbard model close to half filling. A direct comparison of several types of instabilities can be found in Ref. \cite{honerkamp2}.

\section{Current carried by the quasiparticle and the Drude weight}

According to Landau-Fermi liquid theory the current $\vec{J}_{\vec{k}_F}$ carried by the quasiparticle $\vec{k}_F$ is given by\cite{theory-of-FL1}
  
\begin{equation}  
  \label{current}
  \vec{J}_{\vec{k}_F}=\vec{v}_{\vec{k}_F}+ \int \frac{dl_F'}{4\pi^2 v_F'} \;
  f^s(\vec{k}_F,\vec{k}'_F)\, \vec{v}_{\vec{k}'_F}.
\end{equation} 
The first term of (\ref{current}) describes the current of the bare 
quasiparticle and the second one the ``backflow'' of the 
medium around it. 
The Drude weight is connected to the quasiparticle current via \cite{theory-of-FL1} 

\begin{equation}  
  \label{Drude}
  D^{\alpha \beta}=\pi e^2 \int \frac{dl_F}{4\pi^2 v_F} \;
  v^{\alpha}_{\vec{k}_F} J^{\beta}_{\vec{k}_F}.
\end{equation}
With a simple symmetry argument it is easy to show that the off-diagonal
components vanish. This remains true even when a
Pomeranchuk instability breaks the point group symmetry of the 
square lattice, inducing a tetragonal symmetry. The Pomeranchuk instability
merely modifies the diagonal elements 
$D^{aa}$ and $D^{bb}$, which would take different values. 
In this section we will assume that no 
Pomeranchuk instability affects the system, so that 
$D^{aa}$=$D^{bb}=D$.

Note that in a Galilean invariant system
the total current is a constant of motion and does not 
change when the interaction is switched on adiabatically. But as a lattice model the Hubbard model does not belong to this class of systems, 
thus the magnitude of $D$ is related to the strength of the 
interaction $U$. Fig. \ref{currtot} shows the change of the Drude weight
when the Landau interaction function that is used to solve
(\ref{current}) and (\ref{Drude}) is approximated by the ladder and bubble 
summations (\ref{f_ladd_sum_uu}) and (\ref{f_ladd_sum_ud}). The thick lines 
show $D(U)/(e^2 \pi)$ for three different values of the chemical potential $\mu$. With increasing interaction $U$, 
the Drude weight is reduced and goes to zero for a finite
$U$. The reduction occurs for smaller interactions the closer the electron density is to the van Hove filling. 
\begin{figure}[h] 
  \begin{center}
    \includegraphics[width=.4 \textwidth]{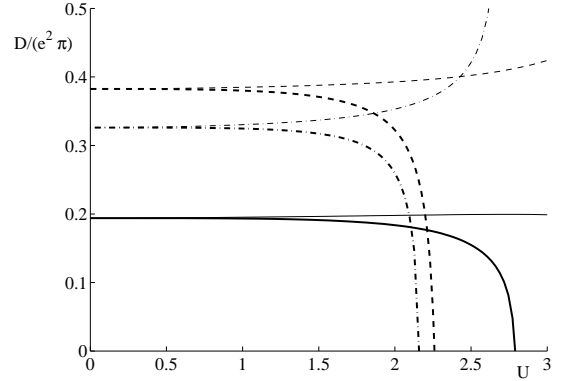} 
    \caption{\footnotesize{The thick lines show $D(U)$ obtained 
        by the ladder summation for three different values of 
        $\mu$: $\mu/t'=-3.33$ dashed
        line, $\mu/t'=-3.99$ dash-dotted line, 
        $\mu/t'=-5.66$ continuous line and T=0.04t. 
        The thin lines show the corresponding results obtained excluding 
        the umklapp process. 
}}
    \label{currtot}
\end{center}
\end{figure}
We note that in our calculation the Drude weight vanishes only at interaction strengths that are larger than the critical value necessary to induce 
the Pomeranchuk instability ($U_p$: $U/t=2.075$ for $\mu=-3.33t$, $U/t=1.91$ for $\mu=-3.99t$ and $U/t=2.6$ for
 $\mu=-5.66t$). Thus the above results have no clear physical meaning for
$U>U_p$. The behavior of the Drude weight for $U < U_p$, however, provides
some useful information: as shown in Fig. \ref{currtot}, its decrease is
strongest at the van Hove filling, $\mu = -4t'$. 
In addition, the results confirm again the important role played by the
umklapp processes. In fact, if they are excluded from the calculation (thin
lines of the Fig. \ref{currtot}), the magnitude of the Drude weight even
increases with the strength of $U$.

It is interesting to determine the scattering processes
responsible for the decrease of $D$. The continuous lines plotted in
Fig. \ref{LADlandau100} show the symmetric Landau interaction function
$f(\theta,\theta')$ for  $\mu/t'=-3.33$ that has been used to solve (\ref{current}) for
different values of $\theta$ and as function of $\theta'$. The thick vertical 
lines mark the different values of $\theta$ and the thin lines are plotted to emphasize the relevant peaks of $f(\theta,\theta')$ for the given $\theta$. 
The insets shows the locations of $\vec{k}_F(\theta)$ (thick line) 
and $\vec{k}'_F(\theta')$ (thin lines) where the peaks in $f(\theta,\theta')$ occur.

\begin{figure}[h] 
  \begin{center}
    \includegraphics[width=.5 \textwidth]{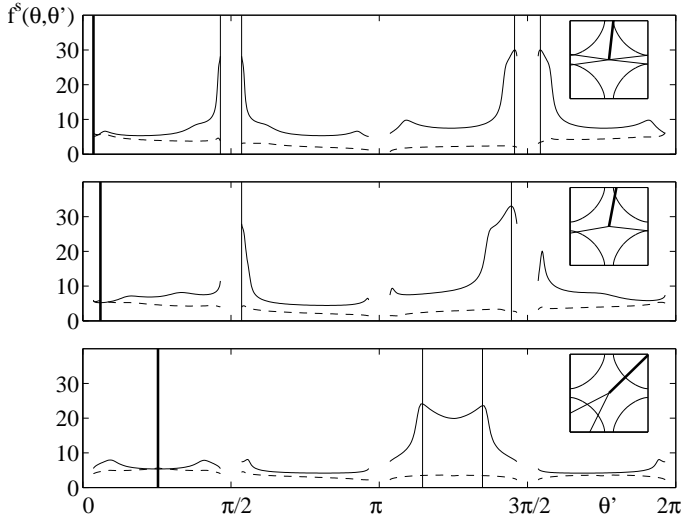} 
    \caption{\footnotesize{The continuous lines show $f(\theta,\theta')$ for 
        $\mu/t'=-3.33$, $U=2.075t$ and $T=0.04t$,
        obtained for three different values of $\theta$, as a function of
        $\theta'$. The thick vertical 
        lines mark the different values of $\theta$ and the thin vertical
        ones are plotted to emphasize the relevant peaks of 
        $f(\theta,\theta')$. The insets show the locations of the Fermi surface points marked by the thick and thin lines. 
        The dashed lines show the corresponding results of 
        $f(\theta,\theta')$ obtained excluding 
        the umklapp process.
}}
    \label{LADlandau100}
\end{center}
\end{figure}  

Fig.\ref{LADlandau100} shows that the peaks in the Landau
function for fillings larger than the van Hove filling are tied to the Fermi surface 
points $\vec{k}_F$, $\vec{k}'_F$ which can be connected by wavevectors close 
to  $(\pi, \pi)$, i.e.  
$\vec{q}_0=\vec{k}_F-\vec{k'}_F \simeq (\pm \pi, \pm \pi)$. These processes
 are enhanced in the ladder and bubble summation of the terms involving
$\chi_{PH}(\vec{q}_0)$ (\ref{chiPH}). For fillings smaller than the 
van Hove filling and $-6.6 <\mu/t'<-4$ the Fermi surface is 
partially nested and the peaks appear when $\vec{q}_0=\vec{k}_F-\vec{k'}_F$
is equal to the corresponding nesting vectors 
(that are smaller than $(\pi,\pi)$). For $\mu/t'< -6.6$ the 
Fermi surface is strictly convex and the 
Drude weight increases with the strength of $U$. 

Without umklapp processes the relevant peaks of $f^s$ disappear, the corresponding Landau interaction (thin dash\-ed lines of
the Fig. \ref{LADlandau100}) causes a ''backflow'' of the medium in the
direction of the bare quasiparticle current implying the increase of
$\vec{J}_{\vec{k}_F}$ for anyvalue of $\vec{k}_F$. Thus the Drude weight will increase.

\section{Conclusions}

We have presented a detailed study of the Landau interaction function $f(\vec{k},\vec{k}')$ and wave-vector resolved compressibilities within Fermi liquid theory for the two-dimensional Hubbard model. This model is often considered as minimal model for the description of high-$T_c$ superconductivity in the copper-oxide materials. 

A less surprising conclusion of this work is that a  Fermi liquid picture
alone can hardly account for the observed anomalies that occur in the underdoped high-$T_c$  materials when the doping is reduced towards half-filling: all 
effects are rather weak and one has to resort to stronger interactions $U$ or
move closer to a magnetic instability (e.g. by ladder summations) to obtain sizable effects such as a significant reduction of the compressibility on parts of the Fermi surface. No rapid decrease of the spin susceptibility reminiscent of spin gap formation\cite{fuseya} can be found. Similar conclusions regarding the high-$T_c$ cuprates were reached by Kim and Coffey\cite{kim} who analyzed the spectral function of the 2D Hubbard model within random phase approximation.
 
On the other hand, our weak coupling analysis provides  
insights into the physical processes which lead to deviations 
from the conventional picture. We find that many of the observed 
tendencies towards partially insulating behavior - as we believe 
an essential ingredient of the underdoped cuprate superconductors - are 
related to elastic umklapp processes. These processes start to affect 
the low energy physics upon reducing the doping when the Fermi surface 
touches the Brillouin zone boundaries. As we know from earlier 
renormalization group studies\cite{furukawa,honerkamp} 
of the same model, strong umklapp processes with momentum transfer $(\pi,
\pi)$ between the saddle point 
regions lead to an interesting mutual reinforcement of antiferromagnetic 
and $d$-wave pairing correlations. Furthermore the renormalization group 
treatment yielded 
a regime with a strong suppression of the charge compressibility near 
the saddle points. A similar effect was observed by Otsuka et 
al.\cite{otsuka} in quantum Monte Carlo calculations for $U=4t$. 
In our Fermi liquid analysis we clearly identify the elastic umklapp 
processes as being responsible for a strong reduction of the Drude weight 
and the angle resolved charge compressibility on parts of the Fermi surface. 
In one-dimensional systems it is well known that elastic umklapp processes 
imply these effects. Our calculations shows that in the 2D Hubbard model 
there are observable tendencies in the same direction. Although our 
perturbative analysis does not allow to conclude
 that umklapp processes are the ultimate cause for the 
pseudogap features at stronger interactions, they will at least contribute 
to a certain extent. Furthermore there 
is no obvious reason why the importance of the umklapp scattering should 
decrease again for stronger interactions.

Our perturbative calculations show tendencies which can be interpreted 
as precursors to stronger reductions in the compressibility and Drude 
weight at stronger interactions, features which appear in some
phenomenological descriptions of the socalled pseudogap region of the 
phase diagram of the underdoped cuprates. In this way they support the
proposal that the critical doping of the cuprates that separates the 
underdoped and overdoped regimes \cite{tallon}, is determined by the
appearance of elastic umklapp scattering with momentum transfer 
$(\pi,\pi)$ connecting the Fermi surface regions near to the saddle
points.

We thank M. Sigrist and M. Troyer for helpful discussions. We acknowledge
financial support by Swiss National Foundation. C.H. also acknowledges
financial support by the German Science Foundation (DFG). Computations were 
carried out on the Beowulfcluster Asgard at ETHZ.

\end{document}